\documentstyle[12pt]{article}

\newcommand{\be}{\begin{equation}}
\newcommand{\ee}{\end{equation}}
\newcommand{\bea}{\begin{array}}
\newcommand{\ea}{\end{array}}
\newcommand{\beqa}{\begin{eqnarray}}
\newcommand{\eeqa}{\end{eqnarray}}
\newcommand{\bean}{\begin{eqnarray*}}
\newcommand{\eean}{\end{eqnarray*}}

\newcommand{\del}{\partial}

\def \bA {{\bar A}}
\def \bdel {{\bar \partial}}
\def \d {{\delta}}

\def \del {{\partial}}
\def \bD {{\bar D}}
\def \bz {{\bar z}}
\def \A {{\cal A}}

\def \G {{\cal G}}
\def \H {{\cal H}}

\def \la {{\langle}}

\def \ra {{\rangle}}
\def \Tr {{\rm Tr}}
\def \vf {{\varphi}}

\def \vx {{\vec x}}
\def \by {{\bar y}}
\def \S {{\cal S}}

\newcommand{\no}{\nonumber}

\begin{document}

\rightline{CCNY-HEP 03/6}
\begin{center}
{\Large \bf Yang-Mills theory, $2+1$ and $3+1$ dimensions
\footnote{Talk at ``Space-time
and
Fundamental Interactions:  Quantum Aspects'', conference
in honor of
A.P. Balachandran's $65$th birthday, Vietri sul Mare, Salerno, Italy,
May 2003; to be published in Mod. Phys. Lett. }
}
\vskip .2in
{V.P. NAIR}
\vskip .1in
{\it Physics Department, City College of the CUNY\\
New York, NY 10031, USA\\
vpn@sci.ccny.cuny.edu}

\end{center}

\centerline{\bf Abstract}

I review the analysis of (2+1)-dimensional Yang-Mills ($YM_{2+1})$ 
theory via the use 
of gauge-invariant matrix variables.
The vacuum wavefunction, 
string tension,
the propagator mass for gluons, its relation to 
the magnetic mass for $YM_{3+1}$ at nonzero temperature
and the extension of our analysis to the
Yang-Mills-Chern-Simons theory are discussed. 
A possible extension to $3+1$ dimensions is also briefly considered.


\section{Introduction}	

In this talk I shall review a Hamiltonian approach to Yang-Mills theory in 
two spatial dimensions
($YM_{2+1}$), on which I have been working for the last few years.
In this analysis, it is possible to carry out some nonperturbative calculations
to the extent that results on mass gap and string tension can be compared with
lattice simulations of the theory. I shall also discuss recent attempts to extend this to
$3+1$ dimensions. The work I
shall report on was done in collaboration with  D. Karabali and
Chanju Kim \cite{1,2,3} and more recently with A. Yelnikov \cite{NY}.

Let me begin with a few general comments.
The $2+1$ dimensional case is interesting
because it 
can be a model simple enough to analyze mathematically 
and yet nontrivial
enough to teach us some lessons about (3+1)-dimensional
$YM$ theories. Another reason to study $YM_{2+1}$ is its relevance to
magnetic screening in $YM_{3+1}$ at high temperature. Gauge theories at finite temperature
have worse infrared problems than at zero temperature due to the divergent nature of the
Bose distribution for low energy modes. A dynamically generated Debye-type screening mass
will eliminate some of these, but we need a magnetic screening mass as well to have a
perturbative expansion which is well defined in the infrared.
The mass gap of the $2+1$ dimensional theory at zero temperature can be interpreted
as the magnetic mass of the $3+1$ dimensional theory at high temperatures,
with a certain relationship between the coupling constants \cite{lingpy}. 
As for the $3+1$ dimensional theory at zero temperature, being part of
the standard model, its importance needs no emphasis.

For $YM_{2+1}$ the following observations will be useful. The
coupling constant
$e^2$ has the dimension of mass and it does not run as the $3+1$-dimensional
coupling does. The dimensionless expansion parameter of the theory is
$k/e^2$ or $e^2/k$, where $k$ is a typical momentum. Thus modes of low momenta
must be treated nonperturbatively, while modes of high momenta can be treated 
perturbatively. There is no simple dimensionless expansion parameter.
$YM_{2+1}$ is perturbatively super-renormalizable, so the ultraviolet  
singularities are well under control. The $3+1$ dimensional theory will require
renormalization and the associated dimensional transmutation.

\section{The $2+1$ dimensional theory}

We shall first discuss $YM_{2+1}$.
Consider 
a gauge theory with group $G=SU(N)$ in the $A_0 =0$ gauge. The gauge potential 
can be written as $A_i = -i t^a A_i ^a$, $i=1,2$, where $t^a$ are hermitian  
$N \times N$-matrices which form a basis of the Lie algebra of $SU(N)$ with 
$[t^a, t^b ] = i f^{abc} t^c,~~{\rm {Tr}} (t^at^b) = {1 \over 2} \delta ^{ab}$. 
The spatial coordinates $x_1 ,x_2$ will be combined into the complex combinations 
$z=x_1 -ix_2,~{\bar z} =x_1+ix_2$ with the corresponding components for the
potential 
$A\equiv A_{z} = {1 \over 2} (A_1 +i A_2), ~~  
{\bar A}\equiv A_{\bar{z}} = 
{1 \over 2} (A_1 -i A_2) = - (A_z)^{\dagger}$. 
The starting point of our analysis is a change of variables given by
\be 
A_z = -\partial_{z} M M^{-1},~~~~~~~~~~~~~ A_{\bar{z}} = M^{\dagger -1} \partial_ 
{\bar{z}} M^{\dagger}   
\label{1}
\ee 
Here $M,~M^\dagger$ are complex matrices in general, not unitary. 
If they are unitary,
the potential is a pure gauge. 
The parametrization (\ref{1})
is standard in many discussions of two-dimensional gauge fields. 
A particular advantage of this parametrization is the way gauge transformations are
realized. A gauge transformation $A_i \rightarrow 
A_i^{(g)} = g^{-1} A_i g + g^{-1} \partial_i g, ~g(x)\in SU(N)$ is obtained 
by the transformation $M\rightarrow M^{(g)}=g M$. The gauge-invariant degrees of freedom
are parametrized by the hermitian matrix $H=M^\dagger M$.
Physical state wavefunctions are functions of $H$.

There will be three basic steps in our analysis, namely, the determination of the inner product 
of the wavefunctions by evaluating
the gauge-invariant volume measure,
rewriting the
Hamiltonian as an operator involving the new variables and finally using these to solve the
Schr\"odinger equation. We will consider these in turn.

The measure of integration over the fields $A, \bA$ is
${d\mu ({\A})/ vol({\G}_*)}$
where
$d\mu ({\cal A})= \prod_{x,a} dA^a (x) d\bA^a (x)$ is the
Euclidean volume element on the space of gauge potentials ${\cal A}$ and  
$vol{\cal G_*}$ is the volume of gauge transformations,
viz., volume of $SU(N)$-valued functions on space.
From (\ref{1}) we see that
$ \delta A= -D(\delta M M^{-1})$,
$\delta \bA = \bD (M^{\dag -1}\delta M^\dag )$
which gives
\be
d\mu ({\cal A}) = (\det D \bD )~d\mu (M, M^\dag )
\label{3}
\ee
where $d\mu (M, M^\dag )$ is the volume for the complex matrices $M, M^\dag$, which is associated
with the metric $ds_M^2~=8 \int {\Tr}(\delta M
M^{-1}~M^{\dagger -1} \delta M^\dagger )$. This is given by the highest order differential form
$dV$ as $d\mu (M, M^\dag )= \prod_x dV(M,M^\dag )$.
Writing out this form and using $M = U\rho$, $H=\rho^2 = M^\dagger M$,
\begin{eqnarray}
d\mu (M, M^{\dagger}) &=& \prod_{x} dV(M, M^{\dag}) 
~ vol (\G _*) 
= vol (\G _*) ~\prod_x d\mu (H)  \label{6}\\
d\mu (H)&=& \epsilon _{a_1...a_n}
(H^{-1}dH)_{a_1} ... (H^{-1}dH)_{a_n}\no\\
vol ({\cal G_*}) &=& \prod_{x} d\mu (U) 
\end{eqnarray}
$d\mu (U)$ is the standard Haar measure for $SU(N)$. 
The volume element or the integration measure for the gauge-invariant configurations 
can now be written as
\begin{eqnarray}
{d\mu ({\A})\over vol({\G}_*)}
&=&{[dA_z dA_{\bar{z}}]\over vol({\G}_*)} \no \\
&=& (\det D_z D_{\bar{z}}) {d\mu  (M,
M^{\dagger})\over vol({\G}_*)} 
=(\det D \bD ) d\mu (H)\label{7}
\end{eqnarray}
The problem is thus reduced to the calculation of the determinant of the
two-dimensional operator $D\bD$. This is well known to be given in terms of the
Wess-Zumino-Witten (WZW) action as \cite{poly}  
\be
(\det D \bD) = \left[ {{\det ' \del \bdel } \over \int d^2 x}
\right] ^{{\rm dim} G} ~ \exp \left[ 2c_A ~\S (H) \right]
\label{8}
\ee
where $c_A \delta^{ab} = f^{amn}f^{bmn}$; it is equal to $N$ for $SU(N)$.
The 
WZW action$\S (H)$ for the hermitian matrix field 
$H$ is given by \cite{witt}
\begin{eqnarray}
{\S} (H) &=& {1 \over {2 \pi}} \int \Tr (\partial H \bar{\partial}
H^{-1}) +{i
\over {12 \pi}} \int \epsilon ^{\mu \nu \alpha} \times\no\\
&&~~~\Tr ( H^{-1} \partial _{\mu} H  H^{-1}
\partial _{\nu}H H^{-1} \partial _{\alpha}H)
 \label{9}
\end{eqnarray}
We can now write the inner product for states $|1\ra$ and
$|2\ra$, represented by the wavefunctions $\Psi_1$ and $\Psi_2$, 
as \cite{GKBN}
\be
\label{inprod}
\la 1 | 2\ra = \int d\mu (H) e^{2c_A ~\S (H)}~~\Psi_1^* \Psi_2 \label{10}
\ee

The next step is the change of variables in the Hamiltonian. However, 
there is some
further simplification we can do before taking up the Hamiltonian. 
The wavefunctions, being gauge-invariant, are functionals of the matrix field
$H$, but actually they can be taken as functionals of the current of the WZW model
(\ref{9}) given by
$J= (c_A/\pi ) \partial_zH~H^{-1}$. 
Equation (\ref{10})
shows that matrix elements of the theory are correlators of the
hermitian WZW model of level number $2c_A$. The properties of the
hermitian model of level number $k+2c_A$ can
be obtained by comparison  with the $SU(N)$-model defined by $e^{k
\S (U)},~ U( x)\in SU(N)$.  The hermitian analogue of the renormalized level $\kappa
= (k+c_A)$ of the
$SU(N)$-model is $-(k+c_A)$. Since the correlators involve only the  renormalized
level
$\kappa$, the correlators of the  hermitian model 
(of level $(k+2c_A)$ ) can
be obtained from the correlators of  the
$SU(N)$-model (of level $k$ ) by the analytic continuation 
$\kappa \rightarrow -\kappa$. For the $SU(N)_k$-model there are the  so-called
integrable representations whose highest weights are limited  by $k$ (spin $\leq k/2$
for $SU(2)$, for example). Correlators involving  the nonintegrable representations
vanish. For the hermitian model the  corresponding statement is that the correlators
involving nonintegrable  representations are infinite.
In our case, $k=0$, and we have only one integrable representation 
corresponding to the
identity operator (and its current algebra descendents). Therefore, for states of
finite norm, it is sufficient to consider $J$ \cite{1,2}.

This means that we can transform the Hamiltonian ${\cal H}= T+V$
to express it in terms of
$J$ and functional derivatives with respect to $J$.
For functionals which can be expanded in powers of
$J$, the kinetic term may worked out by
the chain rule of differentiation, calculating the coefficients of
derivatives with respect to the $J$'s using proper regularization.
We find
\begin{eqnarray}
T&=& {e^2\over 2}\int E^a_iE^a_i = m \Biggl[ \int_u J^a(u) {\d \over \d J^a(u)}
+\int \Omega_{ab} (u,v) 
{\d \over \d J^a(u) }{\d \over \d J^b(v) }\Biggr] \label{13a}\\
V&=&{1\over 2e^2} \int B^aB^a = { \pi \over {m c_A}} \int \bdel J_a (\vx)
\bdel J_a (\vx) \label{13b}
\end{eqnarray}
where $m= e^2c_A/2\pi$ and
\be
\Omega_{ab}(u,v) = {c_A\over \pi^2} {\d_{ab} \over (u-v)^2} ~-~ 
i {f_{abc} J^c (v)\over {\pi (u-v)}}\label{13c}
\ee
The first term in $T$ shows that every power of $J$ in the wavefunction gives
a value $m$ to the energy, suggesting the existence of a mass gap.

The next step is to solve the Schr\"odinger equation.
We will consider the vacuum wavefunction, which
is presumably the simplest to calculate.

For the vacuum state, we take the ansatz
$\Psi_0 = \exp (P)$, where $P$ is taken to be a series in powers of
$J$. 
Substituting this into the Schr\"odinger equation, with the Hamiltonian
given by (\ref{13a}, \ref{13b}), $P$ can be determined term by term.
Upto three powers of the current we find
\begin{eqnarray}
P=-{1 \over {2 e^2}} \int B^a(x_1) K(x_1,x_2) B^a(x_2)
&+& \int  f^{abc}(x_1, x_2, x_3) J^a (x_1) J^b (x_2) J^c(x_3)+ \cdots \no\\
K(x_1,x_2)&=&\left[{ 1 \over  {\bigl( m + 
\sqrt{m^2 - \nabla ^2 } \bigr)} }\right] _{x_1,x_2}
\label{wavfn}
\end{eqnarray}
$f^{abc}(x_1,x_2,x_3)$ is given in
reference [3].
The first term in (\ref{wavfn}) has the correct (perturbative) high
momentum limit, viz.,
\be
\Psi_0 \approx  \exp\Biggl\{ -{1 \over {2 e^2}} \int_{x,y} B^a(x)
\left[{ 1 \over
\sqrt{ - \nabla ^2 } }\right] _{x,y} B^a(y)
+ {\cal O}(3J)\Biggr\}
\label{wavfn2}
\ee
The terms with higher number of $J$'s
can be shown to be small for the low momentum limit and for 
the high momentum limit.

We now use this result to calculate the expectation value of the
Wilson loop operator which is given as
\be
W(C)= \Tr P ~e^{-\oint_C (Adz+\bA d{\bar z})}
=\Tr P ~e^{(\pi /c_A)\oint_C J }
\label{11}
\ee
For the fundamental representation,
its expectation value is given by
\begin{eqnarray}
\la W_F (C) \ra &&= {\rm constant}~~\exp \left[ - \sigma {\cal
A}_C \right]\no\\
{\sqrt{\sigma }}&&= e^2 \sqrt{{N^2-1\over 8\pi}}
\label{tension}
\end{eqnarray}
where ${\cal A}_C$ is the area of the loop $C$. $\sigma$ is the string tension.
This is a prediction of our analysis starting from first principles with no adjustable
parameters. Notice that the dependence on $e^2$ and $N$ is
in agreement 
with large-$N$ expectations, with $\sigma$ depending only on the combination
$e^2N$ as $N\rightarrow \infty$. (The first correction to the large-$N$ limit
is negative, viz., $-(e^2N)/2N^2\sqrt{8\pi}$ which may be interesting in the context
of large-$N$ analyses.)
Formula (\ref{tension}) gives the values $\sqrt{\sigma}/e^2
=0.345, 0.564, 0.772, 0.977$ for $N=2,3,4,5$. 
There are estimates for $\sigma$ based on Monte Carlo simulations of lattice gauge
theory. The results for the gauge  groups $SU(2),~SU(3),
~SU(4)$ and $SU(5)$ are 
$\sqrt{\sigma}/e^2 =$ 0.335, 0.553, 0.758, 0.966 \cite{teper}. We see that
our result agrees with the lattice result to within $\sim 3\%$.

One might wonder at this stage why the result is so good when we have
not included the $3J$- and higher terms in the the wavefunction.
This is basically because the
string tension is determined by large area loops and for these, it is the
long  distance part of the wavefunction which contributes significantly.
In this limit, the $3J$- and higher terms in (\ref{wavfn}) are small compared to the
quadratic term. We expect their contribution to $\sigma$ to be small as well; this is
currently under study.

Another feature of the wavefunction is that
$P$ is nonlocal
when expressed in terms of the magnetic field.
This is essentially due to 
gauge invariance combined with our choice of $A_0=0$; it has recently been shown that
a similar result holds for the Schwinger model \cite{mansfield}.

Some observations on the magnetic mass of $YM_{3+1}$ at finite temperature
can now be made.
The expression (\ref{13a})
shows that for a wavefunction which is just $J^a$, we have the exact
result $T ~J^a = m ~J^a$. 
When the potential term is added, $J^a$ is no longer an exact eigenstate; we find
$(T+V) ~J^a = \sqrt{m^2 -\nabla^2}~ J^a ~+~\cdots$,
showing that the mass value is corrected to the relativistic dispersion relation.
$J^a$ may be considered as the gauge-invariant definition
of the gluon. This result thus suggests a dynamical propagator
mass $m= e^2c_A/2\pi$ for the gluon.
A different way to see this result is as follows.
We can expand the matrix field $J$ in powers of $\vf_a$ which parametrizes $H$, 
so that
$J\simeq (c_A/\pi ) \partial \vf_a t_a$. This is like a perturbation expansion, but
a resummed or improved version of it, where we expand the WZW action in
$\exp(2c_A\S (H))$ but not expand the exponential itself. The Hamiltonian can then be
simplified as
\begin{equation}
\H \simeq {1\over 2} \int_x \left[- {\d ^2 \over {\d \phi _a ^2 (x)}} +
\phi_a (x)  \bigl( m^2 -
\nabla ^2 \bigr)  \phi_a (x)\right]+\cdots
\label{hamil2}
\end{equation}
where $\phi_a (k) \!=\! \sqrt {{c_A k \bar{k} }/ (2 \pi m)} \vf _a (k)$, in momentum space.
This result is obtained by expanding
the currents and also absorbing the
WZW-action part of the measure into the wavefunctions, i.e.,
the operator (\ref{hamil2}) acts on 
${\tilde \Psi} =e^{c_AS(H)}\Psi$. 
The above equation shows that the propagating particles in the
perturbative regime, where the power series expansion of the current is
appropriate, have a mass $m=e^2c_A/2\pi$. 
This value can therefore be identified as the magnetic
mass of the gluons as given by this  nonperturbative analysis.

For $SU(2)$ our result is $m\approx 0.32 e^2$.  
Gauge-invariant resummations of perturbation theory have given the values
$0.38e^2$ \cite{AN} and $0.28e^2$ \cite{owe1}. 
Lattice estimates of this mass are
$0.31e^2$ to $0.40e^2$ (as a common factor mass for glueballs \cite{owe2}) and
$ 0.44e^2$ to
$0.46e^2$ in different gauges \cite{karsch}. 
An unambiguous lattice evaluation of this would require a gauge-invariant
definition a gluon operator and its correlator. Philipsen has recently
given a definition of the ``gluon" on the lattice. Preliminary
numerical estimates
then give the mass as $0.37e^2$ \cite{owe3}.

Let us now turn to excited states.
Eventhough $J$ is useful as  a description of the gluon, it is not
a physical state. This is because of an ambiguity in 
our parametrization (\ref{1}). Notice that
the matrices $M$ and $M{\bar V}(\bz )$ both give the same $A, \bA$, where
${\bar V}(\bz )$ only depends on $\bz$ and not $z$.  Since we have the same potentials,
physical results must be insensitive to this redundancy in the choice of $M$;
in other words, physical wavefunctions must be invariant under $M\rightarrow
M {\bar V}(\bz )$. $J$ is not invariant; we need at least two $J$'s to form
an invariant combination. An example is
\be
\Psi _2 = \int _{x,y} f(x,y) ~
\Bigl[ \bdel J_a (x) \bigl(
H(x,\by) H^{-1} (y,
\by) \bigr) _{ab} \bdel J_b (y)  \Bigr]\label{excited1}
\ee
This is not quite an eigenstate of the Hamiltonian.
Neglecting certain ${\cal O}(J^3)$-terms, one can show that this is an approximate
eigenstate of eigenvalue $E$ if
\be
\Biggl[\sqrt{m^2-\nabla_1^2}~ +\sqrt{m^2-\nabla_2^2}
+\log (\vert x-y\vert^2 /\lambda)\Biggr] f(x,y)= E ~f(x,y)
\label{glueball}
\ee
where $\lambda$ is a scale parameter \cite{2}.
An {\it a posteriori} justification of this would
require that the size of the bound state be not too large on the scale of $1/m$. This does not seem
to be the case, so, at least for now, all we can say is that the mass of $\Psi_2 \geq 2m$; 
see  however \cite{2}.

\section{The Yang-Mills-Chern-Simons Theory}

We have extended our analysis to the Yang-Mills-Chern-Simons theory
by adding a level $k$ Chern-Simons term to the action, 
which gives a perturbative mass
$e^2/4\pi$ to the gluon \cite{YMCS}. The inner product now becomes
\be
\la 1 | 2\ra = \int d\mu (H) e^{(k+2c_A)~\S (H)}~~\Psi_1^* \Psi_2 
\label{ymcs1}
\ee
The earlier conformal field theory argument shows that there are now new integrable
operators. These lead to screening behaviour for the Wilson loop operator
rather than confinement, as expected. The kinetic energy term becomes
\be
T= {e^2\over 4\pi}(k+2c_A) \int_u J^a(u) {\d \over \d J^a(u)}\no\\
+{e^2c_A\over 2\pi}\int \Omega_{ab} (u,v) 
{\d \over \d J^a(u) }{\d \over \d J^b(v) }
\label{ymcs2}
\ee
The gluon mass is now $(k+2c_A)e^2/4\pi$ reflecting dynamical 
mass generation as well
by nonperturbative effects. The vacuum state and a number of excited states
have also been constructed in this case \cite{YMCS}.

\section{Towards the $3+1$ dimensional theory}

The success of our analysis for $YM_{2+1}$ depended largely on 
the use of the parametrization (\ref{1}). This facilitated the 
calculation of the inner product and
the change of variables in the Hamiltonian. So, as a first step
in attempting a generalization to $YM_{3+1}$, we can try to get 
a suitable parametrization for a gauge potential $A^a_i$,
$i=1,2,3$, on ${\bf R}^3$. ($A_0$ can be taken zero as before.)
Recently, A. Yelnikov and I have found a parametrization which may be
appropriate \cite{NY}. For an $U(N)$ gauge field, the matrix $M$ in terms of which
the potentials can be parametrized must take values in
$U(2N, {\bf C})$; this leads to more redundancy than in the
$2+1$ dimensional case and will require
further constraints. Specifically, let $t^a$ be a basis of 
the Lie lagebra
of $U(N)$ (considered as $N\times N$ matrices), 
and let $\sigma_i$ denote the Pauli matrices.
Then a basis for $U(2N, {\bf C})$ is given by $\{ {\bf 1}\otimes t^a, 
{\bf 1}\otimes it^a,
\sigma_i\otimes t^a, \sigma_i\otimes it^a\}$. One can then show that
for every gauge potential $A^a_i$, one can construct a
$U(2N, {\bf C})$-valued matrix $M$ by solving the equation
\be
\sigma\cdot A = - \sigma\cdot \partial M~M^{-1}\label{4ym1}
\ee
Conversely, if an arbitrary $M\in U(2N, {\bf C})$ is given,
one can obtain a gauge potential $A^a_i$ from the above equation,
provided $M$ obeys the further constraints
\begin{eqnarray}
\Tr (t^a \sigma\cdot \del M ~M^{-1} - {\rm h.c.}) &=&0\no\\
\del_i ( M^\dagger \sigma_i M)&=& 0\label{4ym2}
\end{eqnarray}
These results hold for a neighbourhood of the flat potential
in the space of potentials.
If we expand $M$ as $M\approx 1 + it^a \varphi^a +i\sigma_kt^a
\Theta^a_k +\cdots$, it is easy to see that these reproduce 
known Abelian gauge field parametrizations.
In an $N\times N$ splitting, we may write
\be
M = \left( \matrix{U&0\cr 0&U\cr}\right) ~N\label{4ym3}
\ee
where $N$ can be obtained as an algebraic function of the 
gauge-invariant quantities $H =M^\dagger M$ and $W_i =M^\dagger \sigma_i M$.
$U$ can be removed by a gauge choice. The invariant measure of 
integration is then given by
\beqa
{d\mu ({\cal A})\over vol({\cal G_*})}
&=& d\mu (H, W) \exp (\Gamma + {\bar \Gamma}) \no\\
&&~~~~~~~~~\delta [ \sigma\cdot \del N N^{-1}
+{\rm h.c.}]~\delta [ \Tr t^a \sigma\cdot \del N N^{-1} - {\rm h.c.}]
\label{4ym4}
\eeqa
$d\mu (H,W)$ is the product over the spatial points of the volume
on $U(2N, {\bf C})/U(N)$ and can be explicitly given \cite{NY}.
Unlike the lower dimensional case, $\Gamma$ can only
be calculated in powers of $F_{ij}$; to the lowest order it is given by
\be
\Gamma +{\bar \Gamma} = -{c_A \over 128} \int F^a_{ij} \left[
{1\over \sqrt{- (\del +A )^2}}\right]^{ab}
F^b_{ij}+\cdots \label{4ym5}
\ee

We are now trying to work out the Hamiltonian in these variables.
The hope is that the use of $\Gamma +{\bar \Gamma}$, upto the order
calculated, will suffice to show that the wavefunctions should contain
a term like $\exp ( - \int F^2 /8\mu )$ with a dimensional parameter
$\mu$ (related to the $\Lambda$ parameter). In this case, by comparison with 
$YM_{2+1}$, one can conclude that the string tension should be
$\sqrt{\sigma} = \mu \sqrt{(N^2-1)/8\pi}$. This is under investigation.

\section*{Acknowledgments}
This work was supported in part by NSF grant number PHY-0070883
and by a PSC-CUNY grant.


\section*{References}

\vspace*{6pt}

\begin{thebibliography}{0}

\bibitem{1}
D. Karabali and V.P. Nair, {\it Nucl. Phys.} {\bf B464} (1996) 135; {\it Phys.
Lett.} {\bf B379} (1996) 141; {\it Int. J. Mod. Phys.} {\bf A12} (1997) 1161.
\bibitem{2}
D. Karabali, Chanju Kim and V.P. Nair, {\it Nucl. Phys.} {\bf B524} (1998) 661.
\bibitem{3} 
D. Karabali, Chanju Kim and V.P. Nair, {\it Phys. Lett.} {\bf B434} (1998) 103.
\bibitem{NY} V.P. Nair and A. Yelnikov, hep-th/0302176.
\bibitem{lingpy}
A.D. Linde, {\it Phys. Lett.} {\bf B96} (1980) 289;
D. Gross, R. Pisarski and L. Yaffe, {\it Rev. Mod. Phys.} {\bf 53} (1981) 43.
\bibitem{poly}
A.M. Polyakov and P.B. Wiegmann, {\it Phys.Lett.} {\bf B141} (1984) 223.
\bibitem{witt}
E. Witten, {\it Commun. Math. Phys.} {\bf 92} (1984) 455;
S.P. Novikov, {\it Usp. Mat. Nauk.} {\bf 37} (1982) 3.
\bibitem{GKBN}
K. Gawedzki and A. Kupiainen, {\it Phys. Lett.} {\bf B215} (1988) 119;
{\it Nucl.Phys.} {\bf B320} (1989) 649; M. Bos and V.P. Nair, 
{\it Int.J.Mod.Phys.} {\bf A5} (1990) 959.
\bibitem{teper}
M. Teper, {\it Phys. Lett.} {\bf B311} (1993) 223; O. Philipsen,
M. Teper and H. Wittig, {\it Nucl.Phys.} {\bf B469} (1996) 445;
M. Teper, hep-lat/9804008 and references therein.
\bibitem{mansfield}
D. Nolland and P. Mansfield, {\it Int. J. Mod. Phys.} {\bf A15} (2000) 429.
\bibitem{AN}
G. Alexanian and V.P. Nair,
{\it Phys.Lett.} {\bf B352} (1995) 435;
\bibitem{owe1}W. Buchm\"uller and O. Philipsen, {\it Nucl.Phys.} {\bf B443}
(1995) 47; R. Jackiw and S.Y. Pi, {\it Phys.Lett.} {\bf B368} (1996) 131;
{\it ibid} {\bf B403} (1997) 297. 
\bibitem{owe2}
W. Buchm\"uller and O. Philipsen, {\it Phys. Lett.} {\bf B397} (1997) 112.
\bibitem{karsch}
F. Karsch {\it et al}, {\it Nucl. Phys.} {\bf B474} (1996) 217;
F. Karsch, M. Oevers and P. Petreczky, {\it Phys. Lett.} {\bf B442} (1998) 291.
\bibitem{owe3}
O. Philipsen, {\it Phys. Lett.} {\bf B521} (2001) 273; talk at {\it Lattice
2001}, hep-lat/0110114.
\bibitem{YMCS}
D. Karabali, Chanju Kim and V.P. Nair, {\it Nucl. Phys.} {\bf B566} (2000) 331.
 
\end{thebibliography}
\end{document}